\documentclass[aps,prl,twocolumn,superscriptaddress,nofootinbib,floatfix]{revtex4-1}

\usepackage{amsmath,amssymb,graphicx,xcolor}
\usepackage[colorlinks=true,linkcolor=blue,citecolor=blue,urlcolor=blue]{hyperref}

\newcommand{\ER}{E_R}
\newcommand{\keV}{\,{\rm keV}}
\newcommand{\MeV}{\,{\rm MeV}}
\newcommand{\GeV}{\,{\rm GeV}}
\newcommand{\TeV}{\,{\rm TeV}}
\newcommand{\Lam}{\Lambda}
\newcommand{\td}{{\rm d}}

\newcommand{\head}[1]{{\noindent\textbf{#1}} --- \ignorespaces} 

\newcommand{\bea}{\begin{equation}\begin{aligned}} 
\newcommand{\eea}{\end{aligned}\end{equation}}
\newcommand{\be}{\begin{equation}}
\newcommand{\ee}{\end{equation}}

\begin{document}

\title{Boosted dark particles and the LZ nuclear recoil event}

\author{Kristjan Kannike}
\affiliation{National Institute of Chemical Physics and Biophysics, R\"avala 10, Tallinn 10143, Estonia}
\author{Martti Raidal}
\affiliation{National Institute of Chemical Physics and Biophysics, R\"avala 10, Tallinn 10143, Estonia}
\author{Alessandro Strumia}
\affiliation{Dipartimento di Fisica dell'Universit\`a di Pisa, Italia}

\date{\today}

\begin{abstract}
The LZ experiment reported one anomalous event with nuclear recoil energy of 248~keV. Galactic halo dark matter (DM) cannot produce such a large energy unless heavier than 74~GeV, and heavy DM with spin-independent interactions would have produced unseen recoils at lower energy. Inelastic DM has been proposed as a way out. We consider, instead, an elastic collision of a dark sector particle with a momentum of at least 123~MeV and mass in the range 1-74~GeV as deuteron break-up in SNO disfavors sub-GeV masses. The absence of accompanying lower-energy recoils favours interactions whose rate grows with the momentum transfer, such as a pseudoscalar-pseudoscalar nucleon operator. This implies no leading-order signal in argon, an annual modulation below 1\%, and recoils in spinful light nuclei where inelastic halo DM gives nothing. 
\end{abstract}

\maketitle

\head{Introduction}
Direct detection experiments look for nuclei hit by the dark matter (DM) particles of the Galactic halo. DM particles move at a velocity $v\sim 10^{-3}$ and typically produce nuclear recoils of tens of keV, so the searches concentrate on that energy region. Decades of searches did not find a DM signal. The LZ collaboration has now extended its search to nuclear recoils up to $270\keV$ and in $2.84$ ton-years of data found one candidate at $\ER = 248\pm 23\,({\rm stat})\pm 23\,({\rm sys})\keV$~\cite{LZ:2026axp}. The event is a single-scatter nuclear recoil deep inside the fiducial volume. Its energy falls near the top of the search window, where known backgrounds are expected to give $0.011$ events. LZ reports a significance of $2.6\sigma$ after the look-elsewhere correction. 
 
The event is not statistically significant evidence for new physics, but motivates consideration of alternative interpretations. Its energy is several times larger than the recoils DM searches were built for. A DM particle of mass $m$ transfers to a xenon nucleus of mass $m_{\rm Xe}$ at most $\ER^{\max} = 2\mu^2 v^2/m_{\rm Xe}$, where $\mu = m\, m_{\rm Xe}/(m + m_{\rm Xe})$ is their reduced mass. This reaches $\ER^{\max} = 248\keV$ only for DM heavier than about $m = 74\GeV$. Such a DM with the usual spin-independent interaction would have produced hundreds of unseen recoils below $100\keV$. Most interpretations proposed so far~\cite{Su:2026rwz,Fan:2026kxx,Freese:2026sga,Wu:2026nhi,Yin:2026jnn,Nomura:2026qyq,DiMauro:2026ldr,Visinelli:2026kgt,Yamashita:2026ump,Smirnov:2026aqk,Du:2026guj,Rodd:2026tyn,McCabe:2026crm,Chattopadhyay:2026ryw} invoke inelastic scattering $\chi N\to \chi^* N$ into a heavier state~\cite{TuckerSmith:2001hy,Bramante:2016rdh}, with a mass splitting $\delta \approx 300-370\keV$ suppressing low-energy recoils. A thermal Higgsino constitutes one such candidate at $\sim 1\TeV$~\cite{Nagata:2014wma, Graham:2024syw,Fan:2026kxx,Freese:2026sga,Wu:2026nhi,Yin:2026jnn} but is excluded, as it would be captured by the Sun through the same interaction, giving rise to neutrinos not seen by IceCube~\cite{Pospelov:2026ewn,IceCube:2025fcu}. Much heavier DM evades this constraint~\cite{Smirnov:2026aqk}. Other explanations have also begun to appear~\cite{Jeesun:2026vzo,Unwin:2026rdp,Lou:2026idn,Gu:2026vto}. Among these, absorption of a $247\MeV$ fermion~\cite{Dror:2019onn,Lou:2026idn} is excluded by KamLAND. Exothermic DM is another possibility \cite{Dent:2026bji,deLima:2026shq}.

Here we explore an alternative way of achieving it. Instead of inelastic scattering, we assume that there is a fast component of DM or DM-like particles, as we proposed~\cite{Kannike:2020agf} for the XENON1T electron recoil excess~\cite{XENON:2020rca}, a signal that was not confirmed by XENONnT~\cite{XENON:2022ltv}. Such fast particles, moving well above the Galactic escape velocity, can carry enough momentum to produce large recoils. They can be generated by DM decays, semi-annihilations~\cite{Hambye:2008bq,DEramo:2010keq,Belanger:2012vp,Belanger:2014bga}, annihilations~\cite{Agashe:2014yua,Kim:2016zjx,Giudice:2017zke}, cosmic rays~\cite{Bringmann:2018cvk,Ema:2018bih} or gravitational slingshots around black holes~\cite{Acevedo:2026xol,Acevedo:2026tur}. Previously, LZ has set bounds on cosmic ray boosted DM cross section at $3.9\times 10^{-33}~\text{cm}^2$ at 90\% C.L. \cite{LZ:2025iaw} with which our results are consistent. Large energy transfer is indeed a defining feature of the LZ event. The observed recoil energy requires a momentum transfer of at least $123\MeV$, and the absence of lower-energy recoils favors interactions that probe the nucleon spin with an amplitude growing with the momentum transfer.  We build on these two facts and work out their consequences for xenon, argon, deuterium and light nuclei with spin.

\smallskip

\begin{figure}[t]
\centering
\includegraphics[width=\columnwidth]{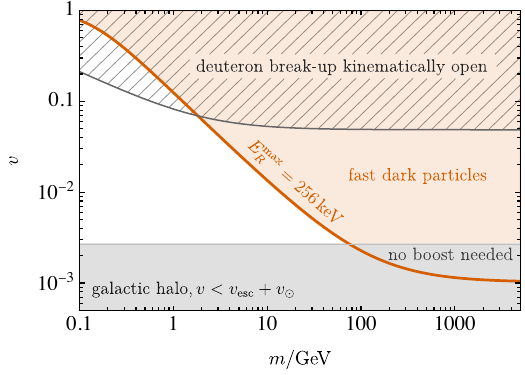}
\caption{\em Velocity of the fast particle reproducing the LZ event as a function of its mass, for the benchmark endpoint $\ER^{\max} = 256\keV$. In the hatched region, the fast particles can break deuterons and SNO strongly disfavors masses below about $1\GeV$. Above $74\GeV$ halo velocities can be enough and no boost is needed.}
\label{fig:mv}
\end{figure}

\head{Kinematics of the fast particle} 
A particle of mass $m$, momentum $p$ and energy $E=\sqrt{p^2+m^2}$ hitting a nucleus $A$ with mass $m_A$ at rest produces recoils up to
\be \label{eq:Emax}
    \ER^{\max} 
    = \frac{2 m_A p^2}{m^2 + m_A^2 + 2 m_A\sqrt{p^2+m^2}} \simeq \frac{2p^2}{m_A}\,,
\ee
where the last expression holds for $m\ll m_A$ and shows a distinctive signature of a light fast particle: $\ER^{\max}$ decreases with the target mass $m_A$. For $m\gg m_A$, the usual approximation $\ER^{\max} \simeq 2 m_A v^2$ is recovered. A recoil of $248\keV$ requires $p\ge p_{\min}\simeq\sqrt{m_{\rm Xe}\ER/2} = 123\MeV$ for $m\ll m_{\rm Xe}$ ($133\MeV$ for $m=10\GeV$, $153\MeV$ for $30\GeV$). The interactions we consider give a recoil spectrum that rises towards its endpoint $\ER^{\max}$, so a single event most likely sits just under it, and $p$ is close to $p_{\min}$. The benchmark $\ER^{\max} = 256\keV$ corresponds to the orange curve in Fig.~\ref{fig:mv}. For $m\gtrsim 74\GeV$, the needed velocity falls below $v_{\rm esc}+v_\odot \approx 800\,{\rm km/s}$~\cite{Baxter:2021pqo} and heavier particles do not need a boost. This is where the grey and orange regions in Fig.~\ref{fig:mv} overlap. In this case, the event is explained by halo DM with momentum-dependent operators~\cite{Fitzpatrick:2012ix,Anand:2013yka,LZ:2023lvz,LZ:2026axp}. We study the complementary mass range, where the LZ event requires particles that cannot be bound to the Galaxy.

\smallskip

\head{A hard spin-dependent interaction}
A boosted momentum $p$ helps to remove the kinematical suppression that contributes to preventing one event at  large recoil $\ER$ not accompanied by many events at lower $E_R$. However, boosting is not enough to entirely solve the issue, because the coherent scattering amplitude $\mathcal{M}$ can be suppressed at large momentum transfer. The expected number of events in LZ is
\be \label{eq:rate}
    N = \Phi\, N_T T \sum_A f_A \int\! \td\ER \epsilon(\ER)\, \frac{\td\sigma_A}{\td\ER}\,,
\ee
where $\Phi$ is the flux of fast particles, $N_T T$ the exposure, $f_A$ the isotopic abundances, and $\epsilon(\ER)$ the LZ nuclear recoil detection efficiency~\cite{LZ:2026axp}, which is $96\%$ in the range $14-250\keV$. For a contact interaction between a fermion $\chi$ and a nucleus $A$ of mass $m_A$, the differential cross section is
\be \label{eq:dsigma}
    \frac{\td\sigma_A}{\td\ER} = \frac{\overline{|\mathcal M|^2}}{32\pi m_A p^2}\,,
\ee
where $\overline{|\mathcal M|^2}$ is the squared spin-averaged amplitude. We consider four Lorentz structures for the $\chi$--nucleon interaction and compare them through $R_{100} \equiv N(\ER < 100\keV)/N(\ER > 200\keV)$, the ratio of low-energy events to detected events above $200\keV$.

\begin{figure}[t]
\centering
\includegraphics[width=\columnwidth]{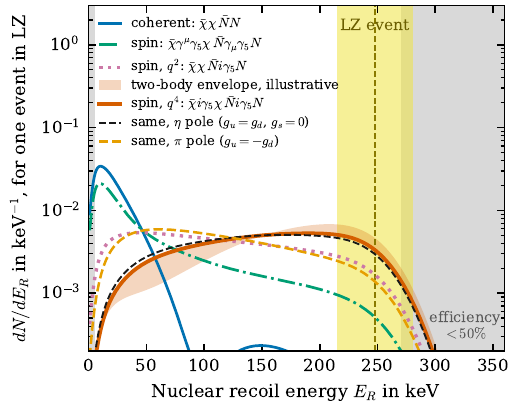}
\caption{\em Recoil spectra expected at LZ for a mono-energetic flux of fast particles with $m=10\,\GeV$ and $\ER^{\max}=256\,\keV$, for four Lorentz structures of the amplitude with $g_p=g_n$. In each case the rate is normalized to one detected event. In the gray shaded regions the LZ detection efficiency is below $50\%$, and the yellow band marks the LZ event. The spectra negligibly depend on $m$ in the range $1-74\,\GeV$.}
\label{fig:spectra} 
\end{figure}

For coherent interactions, such as scalar $\bar\chi\chi\,\bar N N$ or vector $\bar\chi\gamma^\mu\chi\,\bar N\gamma_\mu N$, we have $\overline{|\mathcal M|^2} \propto A^2 F^2(q)$ up to slowly varying factors. In that case, the spectrum follows the nuclear form factor $F^2(q)$~\cite{Helm:1956zz,Lewin:1995rx}, which, at the momentum transfer of LZ, $q\approx\sqrt{2 m_A \ER}=245\MeV$, is suppressed by about $F^2(q)/F^2(0) = 2\times 10^{-4}$ when compared to $q=0$ and lies close to a diffraction zero at $E_R = 280\keV$. As shown in Fig.~\ref{fig:spectra}, the events are then most likely to pile up at lower $\ER$, where none are seen. The coherent operators put $98\%$ of the events below $100\keV$ and give $R_{100} = 370$.

Interactions with the nucleon spin instead have $\overline{|\mathcal M|^2}\propto (m_A/m_n)^2 S(q)$, where $m_n$ is the nucleon mass and $S(q)$ is the spin structure function of the odd isotopes $^{129}$Xe and $^{131}$Xe, which make up $48\%$ of natural xenon and whose spin is carried by neutrons~\cite{Menendez:2012tm,Klos:2013rwa}. Spin responses are not coherent and fall more slowly: in the one-body calculation~\cite{Anand:2013yka} used by LZ, the longitudinal response $S_L(q)$ at $245\MeV$ keeps $1.2\%$ ($^{129}$Xe) and $6.8\%$ ($^{131}$Xe) of its value at $q=0$. Two-body currents, arising from pion exchange between nucleons, modify the spin response and are discussed in Appendix~A. The Lorentz structure of the interaction determines how $\overline{|\mathcal{M}|^2}$ depends on $q$, and hence whether the recoil spectrum peaks near the LZ event or below it.

Fig.~\ref{fig:spectra} also shows, for a mono-energetic flux with $\ER^{\max}=256\keV$, the recoil spectra in LZ for the three spin-dependent structures, normalized to one detected event. The green curve shows the standard spin-dependent interaction given by the axial-vector operator $\bar\chi\gamma^\mu\gamma_5\chi\,\bar N\gamma_\mu\gamma_5 N$. It still gives $R_{100} = 16$, because its amplitude does not grow with $q$. The operator $\bar\chi\chi\,\bar N i\gamma_5 N$, shown by the pink dashed curve, whose nucleon current is the pseudoscalar $\sigma\cdot q$, gives $R_{100} = 3$. 

Each pseudoscalar current adds one power of $q$ into the amplitude, so the pseudoscalar-pseudoscalar operator
\be \label{eq:PP}
    \mathcal{L}
    \!=\!\frac{1}{\Lam^2}\,(\bar\chi\, i\gamma_5\chi)(\bar N\, i\gamma_5 N) \,, 
    \quad
    \overline{|\mathcal M|^2}\!=\!\frac{q^4}{\Lam^4}\frac{m_A^2}{m_n^2} S_L(q) \,,
\ee
has a rate growing as $q^4$. For a nucleus of spin $J$ with proton and neutron couplings $g_{p,n}$, the longitudinal spin response $S_L(q)$ at $q=0$ is $S_L(0) = \frac{4}{3}\frac{J+1}{J}(g_p\langle S_p\rangle + g_n\langle S_n\rangle)^2$. The one-body calculation~\cite{Anand:2013yka} gives $S_L(0) = 0.26$ for $^{129}$Xe and $0.09$ for $^{131}$Xe at $g_p=g_n=1$. Using these, the pseudoscalar-pseudoscalar operator gives $R_{100} = 0.7$, with $21\%$ of the recoils below $100\keV$, $49\%$ between $100$ and $200\keV$ and $30\%$ above $200\keV$. The corresponding spectrum is depicted by the orange curve in Fig.~\ref{fig:spectra}, with the orange band showing the uncertainty from two-body nuclear currents (Appendix~A).

The absence of low-energy events disfavors interactions that give large $R_{100}$. We therefore adopt the pseudoscalar-pseudoscalar operator~\eqref{eq:PP} as our benchmark and fix its isospin structure from the quark-level couplings (Appendix~A). We define the benchmark by the coefficients $g_q$ of $(\bar\chi i\gamma_5\chi)(\bar q i\gamma_5 q)$ at $\mu=2\,\GeV$ in the $\overline{\rm MS}$ scheme, with $g_u=g_d$ and $g_s=0$. At leading chiral order, the nucleon matrix element of $\bar q i\gamma_5 q$ is given by pion and $\eta$ poles~\cite{Bishara:2017pfq}. 
The pion term is proportional to $g_u-g_d$ and problematically falls as $m_\pi^2/(m_\pi^2+q^2)$. 
For isovector couplings ($g_u=-g_d$, $g_p=-g_n$), the $q^4$ growth saturates above $q\approx m_\pi$, i.e.\ above $80\keV$ of recoil, and the spectrum softens to the dashed orange curve of Fig.~\ref{fig:spectra}. For equal couplings, the pion pole cancels and $g_p=g_n$ through the $\eta$ pole, whose factor $m_\eta^2/(m_\eta^2+q^2)$ is $0.83$ at the event, so the spectrum shifts only slightly towards lower $\ER$ (dashed black curve). The pion term is $30$ times the $\eta$ term at $q=245\MeV$. The $\eta$ pole therefore dominates only if $|g_u-g_d|\lesssim 0.03\,|g_u+g_d|$. Couplings proportional to the quark masses, as from a mediator mixing with the Higgs, give $0.37$ instead and leave the pion pole dominant by a factor $11$. 
Isospin breaking in QCD mixes $\pi^0$ and $\eta$ and induces a pion component of about $30\%$ of the $\eta$ term at the event regardless of the couplings. The spectrum stays hard, with $21$--$37\%$ of the events above $200\keV$. The benchmark is therefore the isoscalar operator, $g_p = g_n\equiv g_N$, from equal and mass-independent couplings to $u$ and $d$, whose spectrum is the dashed black curve of Fig.~\ref{fig:spectra}. Xenon sees it through its neutron spin, and fluorine through its proton spin. The contact form holds for a mediator heavier than the momentum transfer, within ten percent for a mass of $1\GeV$ (Appendix~A).\footnote{Another possibility is a CP-odd gluonic source.
This is isoscalar,  so its pion pole vanishes in the exact isospin limit.
The $\eta'$ couples directly via the anomaly, and feeds the $\eta$
through $\eta$--$\eta'$ mixing, while $m_u\neq m_d$ regenerates a pion pole. 
As a result the pion propagator partially overcomes the isospin suppression at low momentum transfer,
so this option does not obviously improve the low-energy tail.}

For the pseudoscalar-pseudoscalar benchmark~\eqref{eq:PP}, the flux required to produce one event in a $2.84$ ton-year exposure is
\be \label{eq:flux}
    \Phi \approx \Big(\frac{\Lam}{\GeV}\Big)^4\, \frac{\tilde\Phi}{{\rm cm}^2\,{\rm s}} \,,
\ee
where $\tilde\Phi = 0.05 - 0.10$ for the contact and the $\eta$-pole couplings, and up to a factor $1.5$ lower if two-body currents enhance the spin response (Appendix~A). At $m=10\GeV$, this is a fraction $\sim 5\times 10^{-8}\,(\Lam/\GeV)^4$ of the total halo DM flux, $\Phi_{\rm DM}\approx\rho_\odot \langle v\rangle/m\approx 10^6 (10\,{\rm GeV}/m) /{\rm cm^2\,s}$. A dark state that constitutes all the DM and decays with a lifetime equal to the age of the universe produces a flux of at most $5\times 10^3\,{\rm cm}^{-2}{\rm s}^{-1}$ for $m=10\GeV$. 
A CP-conserving pseudo-scalar mediator of mass $m_a$ with couplings $g_\chi$ to $\chi$ and $g_N$ to nucleons 
generates at tree level the desired operator with coefficient
$\Lam = m_a/\sqrt{g_\chi g_N}$,\footnote{The mediator can be light enough that loop effects, which generate the coherent spin-independent operator that Eq.~\eqref{eq:PP} avoids, stay small.} and this flux bound translates into $\Lam\lesssim 17\GeV$. For any $\Lambda \gtrsim 1\GeV$, the mean free path of the fast particles in the Earth exceeds its diameter by orders of magnitude, so the flux reaching the detector is not attenuated.

\smallskip

\head{Light nuclei and the lower bound on the mass} 
In neutrino detectors, a fast particle scattering off a nucleon inside the deuteron can break it up. The released neutron is then captured, mimicking the neutral-current solar-neutrino signal~\cite{SNO:2002tuh,SNO:2011hxd}. The break-up $\chi d\to\chi p n$ requires a kinetic energy $\sqrt{p^2+m^2}-m \ge B\,(1+m/m_d)$, where $B=2.22\MeV$ is the deuteron binding energy and $m_d$ its mass. Inverting this gives a threshold velocity $v_{\rm th}(m)$, shown as the hatched region in Fig.~\ref{fig:mv}. For the benchmark momentum $p\approx 127\MeV$, the threshold is crossed only for $m<1.8\GeV$. Above this mass, the fast particle lacks enough kinetic energy to break the deuteron. 

For $m\lesssim m_n$ the momentum transfer to a free nucleon is as large as to xenon, $q\le 2p$, while a free nucleon has no form factor. The cross section on a free nucleon is
\be \label{eq:sigman}
    \sigma_n = \frac{m_n (E_n^{\max})^3}{24\pi p^2 \Lam^4} \stackrel{m\ll m_n}\simeq 5.7\times 10^{-33}\,{\rm cm^2}\,\Big(\frac{\GeV}{\Lam}\Big)^4\,,
\ee
with $E_n^{\max}$ from Eq.~\eqref{eq:Emax}. The product $\Phi\sigma_n$ is independent of $\Lam$ when the flux is set by eq.~\eqref{eq:flux}. Well above the break-up threshold the nucleons are quasi-free, giving about $4\times 10^5$ break-ups per kton-year for $m = 0.3\GeV$ and $3\times 10^4$ for $m=1\GeV$. Sudbury Neutrino Observatory (SNO) measured $5\times 10^3$ neutral-current events per kton-year from $^8$B neutrinos, in agreement with the solar model within $15\%$. At most $10^3$ extra break-ups are allowed. The quasi-free estimate exceeds this by a factor $30$ at $1\GeV$, so SNO strongly disfavors $m\lesssim 1\GeV$ (the range $1-1.8\GeV$ is discussed in Appendix~B). Above $1.8\GeV$, the deuteron cannot break at all. In xenon, nucleon knock-out is forbidden too, while the excitation of the $40$ and $80\keV$ levels of the odd isotopes remains open (Appendix~B).

For $m\gg m_n$, the momentum transferred to a free nucleon is at most $q = 2\mu_n v\simeq 2 m_n p/m$, and the cross section on free nucleons falls as $\sigma_n \propto (m_n/m)^6$, so neutrino detectors lose sensitivity rapidly above $m\sim m_n$. Above $2\GeV$, light nuclei barely constrain the fast particles, and the spin-sensitive fluorine target of PICO-60~\cite{PICO:2019vsc} expects at most $0.05$ events for $g_p=g_n$ in its full exposure, bounding $g_p/g_n < 10$ at $m=2\GeV$ and $< 23$ at $30\GeV$. Free protons in Borexino and KamLAND are insensitive above $1.5\GeV$ (Appendix~B).

\smallskip

\head{Origin of the fast particles} 
As shown above, to explain the LZ candidate event, we need a mono-energetic flux $\Phi\approx 0.06\,(\Lam/\GeV)^4/{\rm cm^2\,s}$ of particles with $m\approx 1-74\GeV$ and $p\approx 125-155\MeV$. The mono-energetic nature of the flux favors two-body decays: a dark state $\psi$ of mass $M$, making up a fraction $f$ of DM, decays as $\psi\to\chi\, S$ into the fast particle $\chi$ and a dark boson $S$ much lighter than $M-m$, giving $\chi$ a momentum $p = (M^2-m^2)/2M$. Since $p\approx 130\MeV \ll M$, the mass splitting is $M - m\approx 130\MeV$, about a pion mass. $\psi$ decays in the galactic halo and $\chi$ arrives at the detector already boosted, with no inelastic interaction required at detection and no tuning of the splitting to the escape velocity. 

The resulting flux of fast $\chi$ particles at Earth is $\Phi = f\Gamma \bar D/M$, with $\bar D = 2.1 \times 10^{22}\GeV/{\rm cm^2}$ being the sky-averaged line-of-sight integral of the DM density~\cite{Navarro:1995iw}, so that Eq.~\eqref{eq:flux} needs
\be \label{eq:fGamma}
    f\,\Gamma \approx 3\times 10^{-23}\,{\rm s}^{-1}\,\Big(\frac{\Lam}{\GeV}\Big)^4\,\frac{M}{10\GeV} \,,
\ee
i.e.\ a lifetime $10^5$ times the age of the universe if $\psi$ is all the DM, or a fraction $f\sim 10^{-5}(\Lam/\GeV)^4$ with a lifetime comparable to the age of the universe. The decays produce only dark particles and convert less than $10^{-5}$ of the DM mass into radiation, so they are invisible to cosmology~\cite{Poulin:2016nat}. The flux follows the DM density so that $19\%$ of it comes from within $30^\circ$ of the Galactic center, $46\%$ from within $60^\circ$. The parent moves with the halo, at $200-800$\,km/s relative to us~\cite{Baxter:2021pqo}, which broadens the line by $m\langle v\rangle/p_0 = 3\%$, $8\%$, $20\%$ for $m = 3$, $10$, $30\GeV$ and softens the spectrum close to the edge: $3\%$ ($12\%$) of the recoils move above $270\keV$ for $m=10$ ($30$)$\GeV$. The motion of the Earth changes the rate by less than $1\%$ throughout the year. Semi-annihilations would select $m=170\MeV$, strongly disfavored by SNO, and gravitational slingshots give broad spectra with too small a fraction at $v\gtrsim 0.01$.

\smallskip

\head{Conclusions}
The LZ experiment reported one anomalous nuclear recoil at $248\keV$. Dark matter bound to the Galaxy cannot produce such a large energy unless heavier than $74\GeV$, and heavy DM with spin-independent interactions would have produced unseen recoils at lower energies. We have proposed, instead, an elastic collision of a fast dark sector particle with a momentum of at least $123\MeV$, produced by decays of a heavier dark state in the galactic halo. The fast particle mass is bounded from below by $m\approx 1\GeV$, as lighter particles are disfavored by SNO through deuteron break-up, and no boost is needed above $74\GeV$. The absence of accompanying lower-energy recoils favors interactions whose rate grows with the momentum transfer. The pseudoscalar-pseudoscalar operator, with a rate $\propto q^4$, gives only $21\%$ of recoils below $100\keV$ and is consistent with the null result of PandaX-4T in its $5-100\keV$ window. For a parent state $\psi$ making up all the dark matter with a lifetime of order the age of the universe, the flux required to explain the LZ event bounds the suppression scale of the pseudoscalar-pseudoscalar operator to $\Lam \lesssim 17\,\GeV$.

If future data confirm the LZ anomaly, new events will discriminate against the coherent and standard spin-dependent halo spectra. The rate does not modulate throughout the year, in contrast to the inelastic interpretations that predict events only in early summer~\cite{Fan:2026kxx,Nomura:2026qyq,DiMauro:2026ldr,McCabe:2026crm}. Argon detectors~\cite{DarkSide-20k:2017zyg,DEAP:2026orr} see no leading-order signal because the abundant stable argon isotopes have $J=0$. In light odd nuclei, the inelastic interpretations yield nothing because the largest splitting they can excite at halo velocities, $\mu v^2/2$, falls below $\delta$~\cite{Bramante:2016rdh}. The fast particles instead give recoils of $^{73}$Ge up to $350-450\keV$ and of $^{19}$F up to $1.5\MeV$ at $m=2\GeV$ and $0.4\MeV$ at $30\GeV$.

\medskip

\begin{acknowledgments}
We thank Hardi Veerm\"ae and Ville Vaskonen for useful comments and feedback on the manuscript. This work was supported by the Estonian Ministry of Education and Research grant TK202 and by the Estonian Research Council grants TARISTU24-TK3, TARISTU24-TK10, PRG3467 and PSG869. We have used Claude and ChatGPT for numerical computations and editing the text.
\end{acknowledgments}

\bibliography{refs}

@article{LZ:2026axp,
    author = "Akerib, D. S. and others",
    collaboration = "LZ",
    title = "{Search for dark matter particle interactions in an extended nuclear recoil energy window with the LUX-ZEPLIN (LZ) experiment}",
    eprint = "2609.02823",
    archivePrefix = "arXiv",
    primaryClass = "hep-ex",
    month = "9",
    year = "2026"
}

@article{Su:2026rwz,
    author = "Su, Liangliang and Yang, Jin Min and Yang, Wen-Na",
    title = "{Inelastic Dark Matter Signature at High Recoil Energy in LUX-ZEPLIN and CRESST}",
    eprint = "2609.01475",
    archivePrefix = "arXiv",
    primaryClass = "hep-ph",
    month = "9",
    year = "2026"
}

@article{Fan:2026kxx,
    author = "Fan, JiJi and Reece, Matthew",
    title = "{Higgsino Above the Sea of Fog}",
    eprint = "2609.01504",
    archivePrefix = "arXiv",
    primaryClass = "hep-ph",
    month = "9",
    year = "2026"
}

@article{Freese:2026sga,
    author = "Freese, Katherine and Theodosopoulos, Dionysios P.",
    title = "{Higgsino Dark Matter Interpretation of the LUX-ZEPLIN 248 keV Nuclear-Recoil Event}",
    eprint = "2609.01583",
    archivePrefix = "arXiv",
    primaryClass = "hep-ph",
    month = "9",
    year = "2026"
}

@article{Wu:2026nhi,
    author = "Wu, Lei and Zhang, Yang and Zhu, Bin",
    title = "{TeV Higgsino Dark Matter from LZ Nuclear Recoil to Fermi-LAT Gamma Rays}",
    eprint = "2609.01590",
    archivePrefix = "arXiv",
    primaryClass = "hep-ph",
    month = "9",
    year = "2026"
}

@article{Yin:2026jnn,
    author = "Yin, Wen",
    title = "{A PQ-Symmetric High-Scale SUSY Interpretation of the LZ High-Energy Recoil}",
    eprint = "2609.01892",
    archivePrefix = "arXiv",
    primaryClass = "hep-ph",
    month = "9",
    year = "2026"
}

@article{BOREXINO:2018ohr,
    author = "Agostini, M. and others",
    collaboration = "BOREXINO",
    title = "{Comprehensive measurement of $pp$-chain solar neutrinos}",
    reportNumber = "FERMILAB-PUB-18-592-ND",
    doi = "10.1038/s41586-018-0624-y",
    journal = "Nature",
    volume = "562",
    number = "7728",
    pages = "505--510",
    year = "2018"
}

@article{Nomura:2026qyq,
    author = "Nomura, Yasunori",
    title = "{Dark Matter as the Z{\_}2 Partner of the Standard Model Higgs Boson}",
    eprint = "2609.02505",
    archivePrefix = "arXiv",
    primaryClass = "hep-ph",
    reportNumber = "RIKEN-iTHEMS-Report-26",
    month = "9",
    year = "2026"
}

@article{DiMauro:2026ldr,
    author = "Di Mauro, Mattia",
    title = "{Dark Matter at the Kinematic Edge: Interpreting the 248 keV LZ Nuclear-Recoil Candidate}",
    eprint = "2609.02608",
    archivePrefix = "arXiv",
    primaryClass = "hep-ph",
    month = "9",
    year = "2026"
}

@article{Visinelli:2026kgt,
    author = "Visinelli, Luca",
    title = "{A Peccei--Quinn Origin for Inelastic Electroweak Dark Matter after LUX-ZEPLIN}",
    eprint = "2609.02807",
    archivePrefix = "arXiv",
    primaryClass = "hep-ph",
    month = "9",
    year = "2026"
}

@article{Yamashita:2026ump,
    author = "Yamashita, Kimiko",
    title = "{Inelastic Dark Photon Dark Matter for the LUX-ZEPLIN High-Recoil Event and the Galactic Halo Gamma-Ray Excess}",
    eprint = "2609.02868",
    archivePrefix = "arXiv",
    primaryClass = "hep-ph",
    month = "9",
    year = "2026"
}

@article{Smirnov:2026aqk,
    author = "Smirnov, Juri and Griffith, Spencer and Beacom, John F.",
    title = "{Inelastic Signatures of Electroweak Dark Matter}",
    eprint = "2609.04144",
    archivePrefix = "arXiv",
    primaryClass = "hep-ph",
    month = "9",
    year = "2026"
}

@article{Du:2026guj,
    author = "Du, Xiaokang and Wang, Fei",
    title = "{TeV Higgsino Interpretation of the LZ High-Recoil Event with Intermediate-Scale Electroweak Gauginos}",
    eprint = "2609.04163",
    archivePrefix = "arXiv",
    primaryClass = "hep-ph",
    month = "9",
    year = "2026"
}

@article{Rodd:2026tyn,
    author = "Rodd, Nicholas L. and Safdi, Benjamin R. and Slatyer, Tracy R. and Xu, Weishuang Linda",
    title = "{Confronting the Higgsino Interpretation of the LZ Event with the High-Energy Sideband}",
    eprint = "2609.04175",
    archivePrefix = "arXiv",
    primaryClass = "hep-ph",
    month = "9",
    year = "2026"
}

@article{McCabe:2026crm,
    author = "McCabe, Christopher",
    title = "{Seasonal dark matter from the LUX-ZEPLIN high-energy event}",
    eprint = "2609.04181",
    archivePrefix = "arXiv",
    primaryClass = "hep-ph",
    month = "9",
    year = "2026"
}

@article{Chattopadhyay:2026ryw,
    author = "Chattopadhyay, Utpal and Das, Debottam and Puri, Rahul and Roy, Joydeep",
    title = "{Sub-TeV Singlino Dark Matter in light from Sagittarius A$^\ast$ and LUX-ZEPLIN Nuclear-Recoil Event}",
    eprint = "2609.02994",
    archivePrefix = "arXiv",
    primaryClass = "hep-ph",
    month = "9",
    year = "2026"
}

@article{TuckerSmith:2001hy,
    author = "Tucker-Smith, David and Weiner, Neal",
    title = "{Inelastic dark matter}",
    eprint = "hep-ph/0101138",
    archivePrefix = "arXiv",
    reportNumber = "UCB-PTH-00-43, LBNL-47234, UW-PT-00-17",
    doi = "10.1103/PhysRevD.64.043502",
    journal = "Phys. Rev. D",
    volume = "64",
    pages = "043502",
    year = "2001"
}

@article{Bramante:2016rdh,
    author = "Bramante, Joseph and Fox, Patrick J. and Kribs, Graham D. and Martin, Adam",
    title = "{Inelastic frontier: Discovering dark matter at high recoil energy}",
    eprint = "1608.02662",
    archivePrefix = "arXiv",
    primaryClass = "hep-ph",
    reportNumber = "FERMILAB-PUB-16-301-T",
    doi = "10.1103/PhysRevD.94.115026",
    journal = "Phys. Rev. D",
    volume = "94",
    number = "11",
    pages = "115026",
    year = "2016"
}

@article{Jeesun:2026vzo,
    author = "Jeesun, Sk and Majumdar, Anirban",
    title = "{Atmospheric neutrino up-scattering explanation of LZ 2026 excess}",
    eprint = "2609.04185",
    archivePrefix = "arXiv",
    primaryClass = "hep-ph",
    month = "9",
    year = "2026"
}

@article{Unwin:2026rdp,
    author = "Unwin, James",
    title = "{Axion Portal Dark Matter and the LUX-ZEPLIN High-Recoil Event}",
    eprint = "2609.04186",
    archivePrefix = "arXiv",
    primaryClass = "hep-ph",
    month = "9",
    year = "2026"
}

@article{Lou:2026idn,
    author = "Lou, Yuanchao and Lu, Chih-Ting",
    title = "{Fermionic Dark Matter Absorption and the High-Energy Event in LUX-ZEPLIN}",
    eprint = "2609.01592",
    archivePrefix = "arXiv",
    primaryClass = "hep-ph",
    month = "9",
    year = "2026"
}

@article{Nagata:2014wma,
    author = "Nagata, Natsumi and Shirai, Satoshi",
    title = "{Higgsino Dark Matter in High-Scale Supersymmetry}",
    eprint = "1410.4549",
    archivePrefix = "arXiv",
    primaryClass = "hep-ph",
    reportNumber = "DESY-14-180, FTPI-MINN-14-37, IPMU14-0320",
    doi = "10.1007/JHEP01(2015)029",
    journal = "JHEP",
    volume = "01",
    pages = "029",
    year = "2015"
}

@article{Graham:2024syw,
    author = "Graham, Peter W. and Ramani, Harikrishnan and Wong, Samuel S. Y.",
    title = "{Enhancing direct detection of Higgsino dark matter}",
    eprint = "2409.07768",
    archivePrefix = "arXiv",
    primaryClass = "hep-ph",
    doi = "10.1103/PhysRevD.111.055030",
    journal = "Phys. Rev. D",
    volume = "111",
    number = "5",
    pages = "055030",
    year = "2025"
}

@article{Pospelov:2026ewn,
    author = "Pospelov, Maxim and Ramani, Harikrishnan",
    title = "{Strong Constraints on Higgsino Dark Matter from Solar Capture}",
    eprint = "2609.02775",
    archivePrefix = "arXiv",
    primaryClass = "hep-ph",
    month = "9",
    year = "2026"
}

@article{IceCube:2025fcu,
    author = "Abbasi, R. and others",
    collaboration = "IceCube",
    title = "{Search for High-Energy Neutrinos From the Sun Using Ten Years of IceCube Data}",
    eprint = "2507.08457",
    archivePrefix = "arXiv",
    primaryClass = "hep-ex",
    month = "7",
    year = "2025"
}

@article{Dror:2019onn,
    author = "Dror, Jeff A. and Elor, Gilly and Mcgehee, Robert",
    title = "{Directly Detecting Signals from Absorption of Fermionic Dark Matter}",
    eprint = "1905.12635",
    archivePrefix = "arXiv",
    primaryClass = "hep-ph",
    doi = "10.1103/PhysRevLett.124.181301",
    journal = "Phys. Rev. Lett.",
    volume = "124",
    number = "18",
    pages = "18",
    year = "2020"
}

@article{Kannike:2020agf,
    author = {Kannike, Kristjan and Raidal, Martti and Veerm{\"a}e, Hardi and Strumia, Alessandro and Teresi, Daniele},
    title = "{Dark Matter and the XENON1T electron recoil excess}",
    eprint = "2006.10735",
    archivePrefix = "arXiv",
    primaryClass = "hep-ph",
    doi = "10.1103/PhysRevD.102.095002",
    journal = "Phys. Rev. D",
    volume = "102",
    number = "9",
    pages = "095002",
    year = "2020"
}

@article{XENON:2020rca,
    author = "Aprile, E. and others",
    collaboration = "XENON",
    title = "{Excess electronic recoil events in XENON1T}",
    eprint = "2006.09721",
    archivePrefix = "arXiv",
    primaryClass = "hep-ex",
    doi = "10.1103/PhysRevD.102.072004",
    journal = "Phys. Rev. D",
    volume = "102",
    number = "7",
    pages = "072004",
    year = "2020"
}

@article{XENON:2022ltv,
    author = "Aprile, E. and others",
    collaboration = "XENON",
    title = "{Search for New Physics in Electronic Recoil Data from XENONnT}",
    eprint = "2207.11330",
    archivePrefix = "arXiv",
    primaryClass = "hep-ex",
    doi = "10.1103/PhysRevLett.129.161805",
    journal = "Phys. Rev. Lett.",
    volume = "129",
    number = "16",
    pages = "161805",
    year = "2022"
}

@article{Hambye:2008bq,
    author = "Hambye, Thomas",
    title = "{Hidden vector dark matter}",
    eprint = "0811.0172",
    archivePrefix = "arXiv",
    primaryClass = "hep-ph",
    reportNumber = "ULB-TH-08-35",
    doi = "10.1088/1126-6708/2009/01/028",
    journal = "JHEP",
    volume = "01",
    pages = "028",
    year = "2009"
}

@article{DEramo:2010keq,
    author = "D'Eramo, Francesco and Thaler, Jesse",
    title = "{Semi-annihilation of Dark Matter}",
    eprint = "1003.5912",
    archivePrefix = "arXiv",
    primaryClass = "hep-ph",
    reportNumber = "MIT-CTP-4136",
    doi = "10.1007/JHEP06(2010)109",
    journal = "JHEP",
    volume = "06",
    pages = "109",
    year = "2010"
}

@article{Belanger:2012vp,
    author = "Belanger, Genevieve and Kannike, Kristjan and Pukhov, Alexander and Raidal, Martti",
    title = "{Impact of semi-annihilations on dark matter phenomenology - an example of $Z_N$ symmetric scalar dark matter}",
    eprint = "1202.2962",
    archivePrefix = "arXiv",
    primaryClass = "hep-ph",
    doi = "10.1088/1475-7516/2012/04/010",
    journal = "JCAP",
    volume = "04",
    pages = "010",
    year = "2012"
}

@article{Belanger:2014bga,
    author = "B{\'e}langer, Genevi{\`e}ve and Kannike, Kristjan and Pukhov, Alexander and Raidal, Martti",
    title = "{Minimal semi-annihilating $\mathbb{Z}_N$ scalar dark matter}",
    eprint = "1403.4960",
    archivePrefix = "arXiv",
    primaryClass = "hep-ph",
    reportNumber = "LAPTH-017-14",
    doi = "10.1088/1475-7516/2014/06/021",
    journal = "JCAP",
    volume = "06",
    pages = "021",
    year = "2014"
}

@article{Agashe:2014yua,
    author = "Agashe, Kaustubh and Cui, Yanou and Necib, Lina and Thaler, Jesse",
    title = "{(In)direct Detection of Boosted Dark Matter}",
    eprint = "1405.7370",
    archivePrefix = "arXiv",
    primaryClass = "hep-ph",
    reportNumber = "MIT-CTP-4538, UMD-PP-014-005",
    doi = "10.1088/1475-7516/2014/10/062",
    journal = "JCAP",
    volume = "10",
    pages = "062",
    year = "2014"
}

@article{Kim:2016zjx,
    author = "Kim, Doojin and Park, Jong-Chul and Shin, Seodong",
    title = "{Dark Matter {\textquotedblleft}Collider{\textquotedblright} from Inelastic Boosted Dark Matter}",
    eprint = "1612.06867",
    archivePrefix = "arXiv",
    primaryClass = "hep-ph",
    reportNumber = "CERN-TH-2016-258",
    doi = "10.1103/PhysRevLett.119.161801",
    journal = "Phys. Rev. Lett.",
    volume = "119",
    number = "16",
    pages = "161801",
    year = "2017"
}

@article{Giudice:2017zke,
    author = "Giudice, Gian F. and Kim, Doojin and Park, Jong-Chul and Shin, Seodong",
    title = "{Inelastic Boosted Dark Matter at Direct Detection Experiments}",
    eprint = "1712.07126",
    archivePrefix = "arXiv",
    primaryClass = "hep-ph",
    reportNumber = "CERN-TH-2017-258, EFI-17-24",
    doi = "10.1016/j.physletb.2018.03.043",
    journal = "Phys. Lett. B",
    volume = "780",
    pages = "543--552",
    year = "2018"
}

@article{Bringmann:2018cvk,
    author = "Bringmann, Torsten and Pospelov, Maxim",
    title = "{Novel direct detection constraints on light dark matter}",
    eprint = "1810.10543",
    archivePrefix = "arXiv",
    primaryClass = "hep-ph",
    doi = "10.1103/PhysRevLett.122.171801",
    journal = "Phys. Rev. Lett.",
    volume = "122",
    number = "17",
    pages = "171801",
    year = "2019"
}

@article{Ema:2018bih,
    author = "Ema, Yohei and Sala, Filippo and Sato, Ryosuke",
    title = "{Light Dark Matter at Neutrino Experiments}",
    eprint = "1811.00520",
    archivePrefix = "arXiv",
    primaryClass = "hep-ph",
    reportNumber = "DESY-18-194",
    doi = "10.1103/PhysRevLett.122.181802",
    journal = "Phys. Rev. Lett.",
    volume = "122",
    number = "18",
    pages = "181802",
    year = "2019"
}

@article{Acevedo:2026xol,
    author = "Acevedo, Javier F. and Ritz, Adam",
    title = "{Binary-boosted Dark Matter}",
    eprint = "2603.08781",
    archivePrefix = "arXiv",
    primaryClass = "hep-ph",
    month = "3",
    year = "2026"
}

@article{Acevedo:2026tur,
    author = "Acevedo, Javier F. and Ritz, Adam",
    title = "{Boosted Dark Matter from Sagittarius A$^\star$}",
    eprint = "2606.30724",
    archivePrefix = "arXiv",
    primaryClass = "hep-ph",
    month = "6",
    year = "2026"
}

@article{Baxter:2021pqo,
    author = "Baxter, D. and others",
    title = "{Recommended conventions for reporting results from direct dark matter searches}",
    eprint = "2105.00599",
    archivePrefix = "arXiv",
    primaryClass = "hep-ex",
    doi = "10.1140/epjc/s10052-021-09655-y",
    journal = "Eur. Phys. J. C",
    volume = "81",
    number = "10",
    pages = "907",
    year = "2021"
}

@article{Fitzpatrick:2012ix,
    author = "Fitzpatrick, A. Liam and Haxton, Wick and Katz, Emanuel and Lubbers, Nicholas and Xu, Yiming",
    title = "{The Effective Field Theory of Dark Matter Direct Detection}",
    eprint = "1203.3542",
    archivePrefix = "arXiv",
    primaryClass = "hep-ph",
    doi = "10.1088/1475-7516/2013/02/004",
    journal = "JCAP",
    volume = "02",
    pages = "004",
    year = "2013"
}

@article{Anand:2013yka,
    author = "Anand, Nikhil and Fitzpatrick, A. Liam and Haxton, W. C.",
    title = "{Weakly interacting massive particle-nucleus elastic scattering response}",
    eprint = "1308.6288",
    archivePrefix = "arXiv",
    primaryClass = "hep-ph",
    doi = "10.1103/PhysRevC.89.065501",
    journal = "Phys. Rev. C",
    volume = "89",
    number = "6",
    pages = "065501",
    year = "2014"
}

@article{Helm:1956zz,
    author = "Helm, Richard H.",
    title = "{Inelastic and Elastic Scattering of 187-Mev Electrons from Selected Even-Even Nuclei}",
    doi = "10.1103/PhysRev.104.1466",
    journal = "Phys. Rev.",
    volume = "104",
    pages = "1466--1475",
    year = "1956"
}

@article{Lewin:1995rx,
    author = "Lewin, J. D. and Smith, P. F.",
    title = "{Review of mathematics, numerical factors, and corrections for dark matter experiments based on elastic nuclear recoil}",
    reportNumber = "RAL-TR-95-024",
    doi = "10.1016/S0927-6505(96)00047-3",
    journal = "Astropart. Phys.",
    volume = "6",
    pages = "87--112",
    year = "1996"
}

@article{Menendez:2012tm,
    author = "Menendez, J. and Gazit, D. and Schwenk, A.",
    title = "{Spin-dependent WIMP scattering off nuclei}",
    eprint = "1208.1094",
    archivePrefix = "arXiv",
    primaryClass = "astro-ph.CO",
    doi = "10.1103/PhysRevD.86.103511",
    journal = "Phys. Rev. D",
    volume = "86",
    pages = "103511",
    year = "2012"
}

@article{Klos:2013rwa,
    author = "Klos, P. and Men{\'e}ndez, J. and Gazit, D. and Schwenk, A.",
    title = "{Large-scale nuclear structure calculations for spin-dependent WIMP scattering with chiral effective field theory currents}",
    eprint = "1304.7684",
    archivePrefix = "arXiv",
    primaryClass = "nucl-th",
    doi = "10.1103/PhysRevD.88.083516",
    journal = "Phys. Rev. D",
    volume = "88",
    number = "8",
    pages = "083516",
    year = "2013",
    note = "[Erratum: Phys.Rev.D 89, 029901 (2014)]"
}

@article{Bishara:2017pfq,
    author = "Bishara, Fady and Brod, Joachim and Grinstein, Benjamin and Zupan, Jure",
    title = "{From quarks to nucleons in dark matter direct detection}",
    eprint = "1707.06998",
    archivePrefix = "arXiv",
    primaryClass = "hep-ph",
    reportNumber = "DO-TH-17-10, OUTP-17-07P, CERN-TH-2017-157",
    doi = "10.1007/JHEP11(2017)059",
    journal = "JHEP",
    volume = "11",
    pages = "059",
    year = "2017"
}

@article{SNO:2002tuh,
    author = "Ahmad, Q. R. and others",
    collaboration = "SNO",
    title = "{Direct evidence for neutrino flavor transformation from neutral current interactions in the Sudbury Neutrino Observatory}",
    eprint = "nucl-ex/0204008",
    archivePrefix = "arXiv",
    doi = "10.1103/PhysRevLett.89.011301",
    journal = "Phys. Rev. Lett.",
    volume = "89",
    pages = "011301",
    year = "2002"
}

@article{SNO:2011hxd,
    author = "Aharmim, B. and others",
    collaboration = "SNO",
    title = "{Combined Analysis of all Three Phases of Solar Neutrino Data from the Sudbury Neutrino Observatory}",
    eprint = "1109.0763",
    archivePrefix = "arXiv",
    primaryClass = "nucl-ex",
    doi = "10.1103/PhysRevC.88.025501",
    journal = "Phys. Rev. C",
    volume = "88",
    pages = "025501",
    year = "2013"
}

@article{PICO:2019vsc,
    author = "Amole, C. and others",
    collaboration = "PICO",
    title = "{Dark Matter Search Results from the Complete Exposure of the PICO-60 C$_3$F$_8$ Bubble Chamber}",
    eprint = "1902.04031",
    archivePrefix = "arXiv",
    primaryClass = "astro-ph.CO",
    reportNumber = "FERMILAB-PUB-19-073-AE-E",
    doi = "10.1103/PhysRevD.100.022001",
    journal = "Phys. Rev. D",
    volume = "100",
    number = "2",
    pages = "022001",
    year = "2019"
}

@article{Poulin:2016nat,
    author = "Poulin, Vivian and Serpico, Pasquale D. and Lesgourgues, Julien",
    title = "{A fresh look at linear cosmological constraints on a decaying dark matter component}",
    eprint = "1606.02073",
    archivePrefix = "arXiv",
    primaryClass = "astro-ph.CO",
    reportNumber = "LAPTH-027-16",
    doi = "10.1088/1475-7516/2016/08/036",
    journal = "JCAP",
    volume = "08",
    pages = "036",
    year = "2016"
}

@article{DarkSide-20k:2017zyg,
    author = "Aalseth, C. E. and others",
    collaboration = "DarkSide-20k",
    title = "{DarkSide-20k: A 20 tonne two-phase LAr TPC for direct dark matter detection at LNGS}",
    eprint = "1707.08145",
    archivePrefix = "arXiv",
    primaryClass = "physics.ins-det",
    reportNumber = "FERMILAB-PUB-17-298-PPD",
    doi = "10.1140/epjp/i2018-11973-4",
    journal = "Eur. Phys. J. Plus",
    volume = "133",
    pages = "131",
    year = "2018"
}

@article{DEAP:2026orr,
    author = "Adhikari, P. and others",
    collaboration = "DEAP",
    title = "{Dark Matter Search with the DEAP-3600 Detector using the Profile Likelihood Ratio Method}",
    eprint = "2603.13965",
    archivePrefix = "arXiv",
    primaryClass = "hep-ex",
    month = "3",
    year = "2026"
}

@article{GAMBITDarkMatterWorkgroup:2017fax,
    author = "Bringmann, Torsten and others",
    collaboration = "GAMBIT Dark Matter Workgroup",
    title = "{DarkBit: A GAMBIT module for computing dark matter observables and likelihoods}",
    eprint = "1705.07920",
    archivePrefix = "arXiv",
    primaryClass = "hep-ph",
    reportNumber = "DESY-17-235, NORDITA-2017-076, gambit-code-2017",
    doi = "10.1140/epjc/s10052-017-5155-4",
    journal = "Eur. Phys. J. C",
    volume = "77",
    number = "12",
    pages = "831",
    year = "2017"
}

@article{GAMBIT:2018eea,
    author = "Athron, Peter and others",
    collaboration = "GAMBIT",
    title = "{Global analyses of Higgs portal singlet dark matter models using GAMBIT}",
    eprint = "1808.10465",
    archivePrefix = "arXiv",
    primaryClass = "hep-ph",
    reportNumber = "ADP-18-22/T1070, COEPP-MN-18-6, DESY-18-141, TTK-18-34, gambit-physics, gambit-physics-2018",
    doi = "10.1140/epjc/s10052-018-6513-6",
    journal = "Eur. Phys. J. C",
    volume = "79",
    number = "1",
    pages = "38",
    year = "2019"
}

@article{Dolan:2014ska,
    author = "Dolan, Matthew J. and Kahlhoefer, Felix and McCabe, Christopher and Schmidt-Hoberg, Kai",
    title = "{A taste of dark matter: Flavour constraints on pseudoscalar mediators}",
    eprint = "1412.5174",
    archivePrefix = "arXiv",
    primaryClass = "hep-ph",
    reportNumber = "DESY-14-238, SLAC-PUB-16179",
    doi = "10.1007/JHEP03(2015)171",
    journal = "JHEP",
    volume = "03",
    pages = "171",
    year = "2015",
    note = "[Erratum: JHEP 07, 103 (2015)]"
}

@article{KamLAND:2013rgu,
    author = "Gando, A. and others",
    collaboration = "KamLAND",
    title = "{Reactor On-Off Antineutrino Measurement with KamLAND}",
    eprint = "1303.4667",
    archivePrefix = "arXiv",
    primaryClass = "hep-ex",
    doi = "10.1103/PhysRevD.88.033001",
    journal = "Phys. Rev. D",
    volume = "88",
    number = "3",
    pages = "033001",
    year = "2013"
}

@article{Baudis:2013bba,
    author = "Baudis, L. and Kessler, G. and Klos, P. and Lang, R. F. and Men{\'e}ndez, J. and Reichard, S. and Schwenk, A.",
    title = "{Signatures of Dark Matter Scattering Inelastically Off Nuclei}",
    eprint = "1309.0825",
    archivePrefix = "arXiv",
    primaryClass = "astro-ph.CO",
    doi = "10.1103/PhysRevD.88.115014",
    journal = "Phys. Rev. D",
    volume = "88",
    number = "11",
    pages = "115014",
    year = "2013"
}

@article{Navarro:1995iw,
    author = "Navarro, Julio F. and Frenk, Carlos S. and White, Simon D. M.",
    title = "{The Structure of cold dark matter halos}",
    eprint = "astro-ph/9508025",
    archivePrefix = "arXiv",
    doi = "10.1086/177173",
    journal = "Astrophys. J.",
    volume = "462",
    pages = "563--575",
    year = "1996"
}

@article{LZ:2023lvz,
    author = "Aalbers, J. and others",
    collaboration = "LZ",
    title = "{First constraints on WIMP-nucleon effective field theory couplings in an extended energy region from LUX-ZEPLIN}",
    eprint = "2312.02030",
    archivePrefix = "arXiv",
    primaryClass = "hep-ex",
    doi = "10.1103/PhysRevD.109.092003",
    journal = "Phys. Rev. D",
    volume = "109",
    number = "9",
    pages = "092003",
    year = "2024"
}

@article{LZ:2025iaw,
    author = "Aalbers, J. and others",
    collaboration = "LZ",
    title = "{New Constraints on Cosmic Ray-Boosted Dark Matter from the LUX-ZEPLIN Experiment}",
    eprint = "2503.18158",
    archivePrefix = "arXiv",
    primaryClass = "hep-ex",
    reportNumber = "FERMILAB-PUB-25-0191-V",
    doi = "10.1103/nr92-jvt3",
    journal = "Phys. Rev. Lett.",
    volume = "134",
    number = "24",
    pages = "241801",
    year = "2025"
}

@article{Dent:2026bji,
    author = "Dent, James B. and Newstead, Jayden L.",
    title = "{Exothermic and Endothermic Inelastic Dark Matter Interpretations at LZ: Sideband Constraints and Future Prospects}",
    eprint = "2609.04673",
    archivePrefix = "arXiv",
    primaryClass = "hep-ph",
    month = "9",
    year = "2026"
}

@article{deLima:2026shq,
    author = "de Lima, Carlos Henrique",
    title = "{Exothermic Dark Matter at LZ}",
    eprint = "2609.05204",
    archivePrefix = "arXiv",
    primaryClass = "hep-ph",
    month = "9",
    year = "2026"
}

@article{Gu:2026vto,
    author = "Gu, Guanhua and Li, Lingfeng and Tang, Shao-Song and Xu, Yongheng",
    title = "{Inelastic from the Other Side: Xenon Excitation Signals in Light of the LZ High-Recoil Event}",
    eprint = "2609.05291",
    archivePrefix = "arXiv",
    primaryClass = "hep-ph",
    month = "9",
    year = "2026"
}

\appendix

\section{Appendix A: Nuclear responses and nucleon matching}
\label{app:nuclear}

The longitudinal response follows from the tabulated one-body $\Sigma''$ responses of~\cite{Anand:2013yka,GAMBITDarkMatterWorkgroup:2017fax,GAMBIT:2018eea} as
\bea
\label{eq:SL}
    S_L = &\frac{4\pi}{2J+1}\bigg[ (g_p+g_n)^2\, W^{00} \\
    &+ (g_p^2-g_n^2) \,(W^{01}+W^{10}) + (g_p-g_n)^2\, W^{11}\bigg] ,
\eea
where $W^{\tau\tau'}(q)$ are the isoscalar ($\tau = 0$) and isovector ($\tau = 1$) components of the $\Sigma''$ response, and the $q\to 0$ limit reproduces the $S_L(0)$ quoted in the main text. The Baldridge--Vary density matrices of~\cite{Anand:2013yka} give $\langle S_p\rangle = 0.007$ and $\langle S_n\rangle = 0.248$ for $^{129}$Xe, and $0.005$ and $0.199$ for $^{131}$Xe, hence $S_L(0) = 0.26$ and $0.09$ at $g_p = g_n = 1$.

Following~\cite{Bishara:2017pfq}, the nucleon matrix element of $m_q\bar q i\gamma_5 q$ is carried by pion and $\eta$ poles, so that for the quark-level operator $(\bar\chi i\gamma_5\chi)\sum_q g_q\,\bar q i\gamma_5 q/\Lam_q^2$ the nucleon couplings are
\bea
\label{eq:gpn}
    g_{p,n}(q^2) = m_N B_0\bigg[&\pm(g_u-g_d)\,\frac{g_A}{m_\pi^2+q^2} \\
    &+ (g_u+g_d-2g_s)\,\frac{\Delta_8}{3\,(m_\eta^2+q^2)}\bigg] \,,
\eea
where the $+ (-)$ sign stands for the proton (neutron), $B_0 = m_\pi^2/(m_u+m_d)\approx 2.8\GeV$, $g_A = 1.27$ and $\Delta_8 = \Delta u+\Delta d-2\Delta s = 0.58$. The nucleon-level scale of Eqs.~\eqref{eq:PP}--\eqref{eq:fGamma} is $\Lam = \Lam_q/\sqrt{g_N(0)}$.

The benchmark $g_u = g_d = 1$, $g_s = 0$ removes the pion term of Eq.~\eqref{eq:gpn} and gives $g_p = g_n$, which is equal to $3.4$ at $q = 0$ and $2.9$ at $245\MeV$. Isovector couplings $g_u = -g_d = 1$ remove the $\eta$ term instead and give $g_p = -g_n$, which is equal to $-350$ and $-85$. Flavour-universal couplings $g_u = g_d = g_s$ remove the octet $\eta$ term as well, leaving the flavour-singlet $\eta'$ and gluonic contributions that the leading-order matching does not fix. Table~\ref{tab:fractions} collects some characteristics of the resulting spectra.

Beyond the one-body response of Eq.~\eqref{eq:SL}, chiral two-body currents enhance the longitudinal spin response. The Klos et al.\ calculation~\cite{Klos:2013rwa} with the GCN5082 interaction gives a factor of $1.8$ enhancement of $S_L$ at $q=0$, falling with $q$ so that the integrated flux required for one event decreases by a factor of $1.5$. We therefore treat the two-body effect as a normalization uncertainty of up to $1.5$ on the flux, with the spectrum given in Table~\ref{tab:fractions} and an operator comparison that changes by less than $30\%$. The flux of Eq.~\eqref{eq:flux} is $0.051-0.079$ for $m=2-30\GeV$ with the contact coupling and $0.064-0.097$ with the $\eta$-pole matching.

Three corrections to the leading-order matching affect the benchmark. The first changes the shape of the spectrum. Isospin breaking mixes $\pi^0$ and $\eta$ with an angle $\epsilon\approx 0.01$ and reintroduces a pion component of relative amplitude
\be
    \epsilon\,\frac{3g_A}{\Delta u+\Delta d-2\Delta s}\, \frac{m_\eta^2+q^2}{m_\pi^2+q^2}\approx 0.3
\ee
Depending on the sign, the spectrum moves within the range given in Table~\ref{tab:fractions}, and the density at $248\keV$ shifts by $20-40\%$ either way.

\begin{table}[t]
\centering
\begin{tabular}{lcccc}
\hline\hline
 & $<100$ & $100-200$ & $>200$ & $R_{100}$ \\
\hline
1-body, contact & $21$ & $49$ & $30$ & $0.7$ \\
1-body, $\eta$ pole & $24$ & $49$ & $27$ & $0.9$ \\
1-body, $\pi$ pole & $46$ & $41$ & $13$ & $3.5$ \\
2-body scaled, contact & $12$ & $48$ & $39$ & $0.3$ \\
$\pi$--$\eta$ mixing & $12-30$ & & $21-37$ & $0.3-1.4$ \\
\hline\hline
\end{tabular}
\caption{\label{tab:fractions} \em Percentage of detected events in the three recoil ranges in keV, and $R_{100}$ defined in the main text, for the variants of the pseudoscalar-pseudoscalar benchmark discussed in this appendix.}
\end{table}

The other two corrections leave the shape unchanged. Non-pole terms enter $F_P^{q/N}$ only at next chiral order, with relative size $m_\eta^2/(4\pi f_\pi)^2\approx 0.2$ on the isoscalar amplitude, since the constant term of~\cite{Bishara:2017pfq} belongs to the CP-odd gluonic form factor rather than to the quark one. The $\eta'$ contributes because our couplings have a flavor-singlet component. We do not include it, and it would add about $m_\eta^2/m_{\eta'}^2\approx 0.3$ of the $\eta$ term if its residue is of the same order. Both are nearly constant below $245\MeV$, the $\eta'$ factor being $0.94$ at the event, so they only rescale $g_N(0)$, which enters through $\Lam = \Lam_q/\sqrt{g_N(0)}$. The shape stays between the contact and $\eta$-pole curves of Fig.~\ref{fig:spectra}, whose fractions differ by at most three percentage points. Only the pion pole changes the shape, which is why $g_u\simeq g_d$ is needed.

A mediator of mass $m_a$ replaces $1/\Lam^2$ with $g_\chi g_N/(q^2+m_a^2)$, so that at $q=245\MeV$ the rate is $0.89$ of the contact value for $m_a=1\GeV$ and $0.99$ for $3\GeV$. We do not address which completions with $g_u = g_d$ survive flavour, beam-dump, and collider bounds~\cite{Dolan:2014ska}.

\medskip

\section{Appendix B: Light targets}
\label{app:light}

{\it Deuterium.} For $m\ll 1\GeV$, the nucleon recoils of Eq.~\eqref{eq:sigman} reach $26\MeV$. Protons contribute equally. Well above the three-body threshold, the nucleons are quasi-free and $\Phi\sigma_n$ does not depend on $\Lam$, giving $9\times 10^5$, $4\times 10^5$, $3\times 10^4$, $7\times 10^3$, $3\times 10^3$ break-ups per kton-year for $m = 0.1$, $0.3$, $1$, $1.5$, $1.8\GeV$ (recoils above the $2.2\MeV$ binding energy, a cut that removes $2\%$ at $1\GeV$). The SNO neutral-current rate, $5\times 10^3$ per kton-year from $^8$B neutrinos~\cite{SNO:2002tuh,SNO:2011hxd}, agrees with the solar model, whose flux is predicted within $15\%$. An extra neutral-current-like signal is therefore limited to about $10^3$ per kton-year, and the excess is $400$, $30$, $7$, $3$ at $0.3$, $1$, $1.5$, $1.8\GeV$. At $1\GeV$ the momentum transfer reaches $122\MeV$, $2.7$ times the deuteron momentum $\sqrt{m_n B} = 46\MeV$, and the two nucleons separate with a relative energy of about $4\MeV$, above the final-state interaction region. The quasi-free treatment should then be good to a factor of order two, so a factor $30$ is beyond what it or the isoscalar structure of the transition can recover. A limit would need the deuteron matrix element of the pseudoscalar operator and the SNO neutron efficiency for neutrons harder than the $^8$B ones, for which it was measured. Above $1.2\GeV$ the margin falls below $15$ while the projectile approaches threshold, so the deuteron wave function and the final-state interactions decide, and we regard $1$--$1.8\GeV$ as indicative.

{\it Free protons.} With $g_p = g_n$, protons in Borexino and KamLAND~\cite{BOREXINO:2018ohr,KamLAND:2013rgu} recoil up to $8$, $5$, $3.5\MeV$ ($3.8$, $2.2$, $1.4\MeV$ electron-equivalent after quenching) for $m = 1$, $1.5$, $2\GeV$, at $2\times 10^4$, $5\times 10^3$, $2\times 10^3$ per kton-year. At $1\GeV$ this exceeds the backgrounds above $3\MeV$, while from $1.5\GeV$ on it sits under the cosmogenic and reactor backgrounds of $10^4-10^5$ per kton-year.

{\it Fluorine.} Eq.~\eqref{eq:PP} reduces to the non-relativistic operator $(\vec S_\chi\!\cdot\!\vec q)(\vec S_N\!\cdot\!\vec q)$ and probes the longitudinal response alone, so we use the one-body $\Sigma''$ response of $^{19}$F from the tabulation used for xenon~\cite{Anand:2013yka}, with $S_L(0) = 4(\langle S_p\rangle+\langle S_n\rangle)^2 = 0.87$ at $g_p=g_n=1$ for $\langle S_p\rangle = 0.475$ and $\langle S_n\rangle = -0.009$. The total axial structure factor of~\cite{Klos:2013rwa} contains the transverse response as well, and its fit turns negative inside the range of momenta that PICO probes ($q$ up to $228\MeV$, $u = 2.1$). The complete PICO-60 exposure~\cite{PICO:2019vsc} combines $1167$ kg\,day at $3.3\keV$ threshold with no candidates and $1404$ kg\,day at $2.45\keV$ with three candidates on $1.0\pm 0.4$ expected. For a hard spectrum, the threshold is immaterial, nucleation is certain, and the acoustic rejection of alphas was calibrated with neutron recoils of MeV energies, so we use the combined $2571$ kg\,day with $3$ observed on $1.0$ expected, allowing $5.7$ signal events at $90\%$ C.L. The benchmark gives $0.05$ events at $m=2\GeV$ and $0.01$ at $30\GeV$. A proton coupling $r_p g_N$ scales the rate as $r_p^2$, so $r_p < 10$ at $2\GeV$ and $23$ at $30\GeV$, up to $20\%$ from the response.

{\it Xenon.} For $m\gtrsim 2\GeV$ nucleon knock-out is closed, while the $39.6$ and $80.2\keV$ levels of $^{129}$Xe and $^{131}$Xe can be excited. The inelastic structure factors of~\cite{Baudis:2013bba}, computed for the axial current, are of the order of the elastic ones at these $q$, but the longitudinal one that Eq.~\eqref{eq:PP} needs is not available, so we quote no yield for such mixed nuclear-electronic deposits.

\end{document}